\documentstyle[aps,multicol]{revtex}
\def\etal{\emph{et al.}}

\input BoxedEPS
\SetepsfEPSFSpecial
\HideDisplacementBoxes

\begin{document}

\draft

\title{Influence of temperature dependent inelastic scattering on the
superconducting proximity effect}

\author{M.\ J.\ Black and V.\ Chandrasekhar}

\address{Department of Physics and Astronomy, Northwestern
University, Evanston, IL 60208}

\date{\today}

\maketitle

\begin{abstract}

We have measured the differential resistance of mesoscopic gold
wires of different lengths connected to an aluminum
superconductor as a function of temperature and voltage.  Our
experimental results differ substantially from theoretical predictions
which assume an infinite temperature independent gap in the superconductor. 
In addition to taking into account the temperature dependence of the
gap, we must also introduce a temperature dependent
inelastic scattering length in order to fit our data.
\end{abstract}


\begin{multicols}{2}
When a superconductor (S) is brought into intimate electrical contact with a
normal metal (N) wire, induced pair correlations caused by Andreev reflection
\cite{Andreev} result in an enhancement of the normal metal diffusion
constant $D_{N}$ \cite{deGennes}.  According to the quasiclassical theory of
superconductivity, this enhancement has a non-monotonic dependence on the energy
$\varepsilon$ of the quasiparticles in the normal metal, with a maximum at a
characteristic energy scale $E_c = \hbar D_N/L^2$, where
$L$ is the length of the wire
\cite{Belzig,Nazarov}.  As a consequence, the differential
resistance of the wire is predicted to have a
non-monotonic dependence as a function of temperature or voltage, with minimum
values at $T_{min} \sim E_c/k_B$ and
$V_{min} \sim E_c/e$ respectively.

A number of groups have reported observing this reentrant behavior in the differential
resistance of normal one-dimensional wires connected to a superconductor.  Charlat
\etal \cite{Charlat} measured the conductance of a Cu loop connected to a
superconducting Al reservoir as a function of temperature, and found good
agreement with quasiclassical predictions \cite{Belzig}.  Similar results were obtained by Petrashov
\etal
\cite{Petrashov1} when they measured magnetoresistance oscillations in Pb-Ag
structures.  However, in other experiments
\cite{denHartog,Toyoda,Courtois,Chien}, $T_{min}$ or $V_{min}$
did not correspond to the
values predicted by theory.  Furthermore, even in a single sample, the voltage and
temperature scales at which the minima in resistance were observed did not agree
($eV_{min} \ll k_B T_{min}$) \cite{Chien}. 

In all these experiments, many simplifying
assumptions were made to make the calculations tractable.  For example, it was
typically assumed that the gap $\Delta$ in the superconductor was temperature
independent and much larger than the energy
$\varepsilon$ of the quasiparticles, an assumption which is clearly not valid in the experimental
regime of interest for those experiments which used Al as the superconductor.  More recently, some
groups have attempted to improve the agreement between theory and experiment by taking these
factors into account \cite{Petrashov2}, but the results are still not satisfactory,
especially near the superconducting transition temperature
$T_c$ of the superconductor.

In order to clarify these issues and provide a quantitative test of the
quasiclassical theory of the proximity effect in mesoscopic samples, we set out to measure
the length dependence of the proximity effect in diffusive Au wires.  The samples
were designed to correspond to the simplest geometry analyzed theoretically by
Nazarov and Stoof \cite{Nazarov}: a single normal Au wire connected on one end to a
superconducting Al reservoir, and to a normal Au reservoir on the other end.  For
wires in which the electron phase coherence length
$L_{\phi}$ is longer than the length $L$ of the wire (as is the case in our
samples), Nazarov and Stoof predict that $V_{min}\sim 5E_c/e$ and $T_{min}\sim
5E_c/k_B$.  However, in our samples, we find
that the situation is much more complicated.  $T_{min}$ is approximately a factor
of 5 less than predicted by Nazarov and Stoof.  In addition, we find that
$V_{min}$ is typically much less than $k_B T_{min}/e$ in the same sample due to heating of
the quasiparticles in the wires by the dc bias, in spite of the fact that $L$ is much
smaller than $L_{\phi}$.  In order to fit our data within
the framework of the quasiclassical theory, we need to take into account the
temperature dependence of
$\Delta$, as well as the temperature dependence of the inelastic scattering length of the
quasiparticles in the normal metal.

The Au/Al samples for this experiment were fabricated by standard electron-beam
lithography techniques onto oxidized Si substrates.  Figure 1 shows a scanning
electron micrograph of one of our samples.  In consists of five Au wires of
length $L$ ranging from $\sim 0.75$-1.5 $\mu$m.  Each wire is connected to a Au
reservoir on one end and a large superconducting Al reservoir on the other.  In
order to minimize variations in the transparency of the NS interface barriers, all
wires are connected to the same Al reservoir.  The Au layer was patterned and
deposited first; following a second level of lithography, the Al film was
deposited immediately after cleaning the Au layer with an oxygen plasma etch.  This cleaning step
guaranteed a transparent NS interface, as evidenced by a consistent interfacial resistance of less than
0.1
$\Omega$.  Unlike most previous experiments, each Au wire also has two voltage probes in order to
enable us to make four-terminal differential resistance measurements of the proximity wire 
\begin{figure}[p]
\begin{center}
\BoxedEPSF{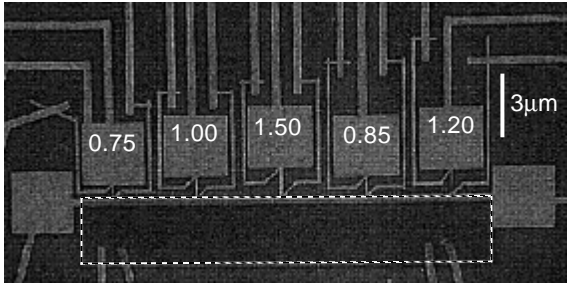 scaled 750}
\end{center}
\caption{Scanning electron micrograph of our device structure.  The
light colored areas are 50 nm thick Au, and the dotted line denotes the
edge of the 80 nm thick Al contact.}
\label{Fig1}
\end{figure}
alone
with no direct contribution from the Al reservoir.

The samples were measured in
a dilution refrigerator using standard four-terminal techniques, with ac currents
small enough to avoid self-heating.  A total of six sample sets were measured: 
three in the geometry of Fig. 1, and three in a similar but slightly different
geometry, with the length of the wires ranging from 0.5-3.0
$\mu$m.  We present here data from three wires in the set shown in Fig. 1; data
from wires in other sets showed similar behavior.  For this set, other relevant
sample parameters are as follows:  Au film thickness 50 nm, Au wire width 120 nm,
Al film thickness 80 nm, Au diffusion constant $D_N$ = 300 cm$^2$/sec, Au thermal
diffusion length $L_T =
\sqrt{\hbar D/k_B T} = 0.48$ $\mu$m at 1 K, and Al superconducting transition
temperature $T_c=1.20$ K.  The electron phase coherence length was estimated
to be 3.7 $\mu$m at 27 mK from weak localization measurements on a co-evaporated Au
wire.

Figure 2(a) shows the zero-bias differential resistance $R(T)$ of three Au
wires of length $L$=1.02, 1.23 and 1.47 $\mu$m from a single set normalized to their
normal state values $R_N$ as a function of
$T$ \cite{samplenote}.  All three wires show
reentrant behavior, with the minimum in resistance occuring at a lower temperature
$T_{min}$ for the longer wires.  It is immediately apparent that
$T_{min}$ does not scale as $1/L^2$.  Furthermore, the observed value of $T_{min}$
is much smaller than the value of $5E_c/k_B$ predicted by Nazarov and Stoof.  For
example, for the $L$=1.47
$\mu$m wire, $5E_c/k_B$=0.534 K, whereas $T_{min} \sim 0.148$ K.  In the case of the dc
voltage bias dependence, which is shown in Fig. 2(b), the situation is even more
complicated.  $V_{min}$ does not show any systematic dependence on $L$:  $V_{min}$
for the 1.02 $\mu$m and 1.47 $\mu$m wires are approximately the same, while $V_{min}$
for the 1.23 $\mu$m wire has a smaller value.  For all wires, $V_{min}$ is again
much less than the theoretical prediction of $5E_c/e$.

The non-systematic dependence of $V_{min}$ on $L$ is due to the increase in the effective
temperature of the quasiparticles in the Au wires due to heating by the dc bias \cite{heatingnote}. 
This can be shown directly by measuring the differential resistance of one Au wire as
a function of dc voltage bias while simultaneously measuring the differential
resistance of a second Au wire adjacent to the first.  The 
\begin{figure}[p]
\begin{center}
\BoxedEPSF{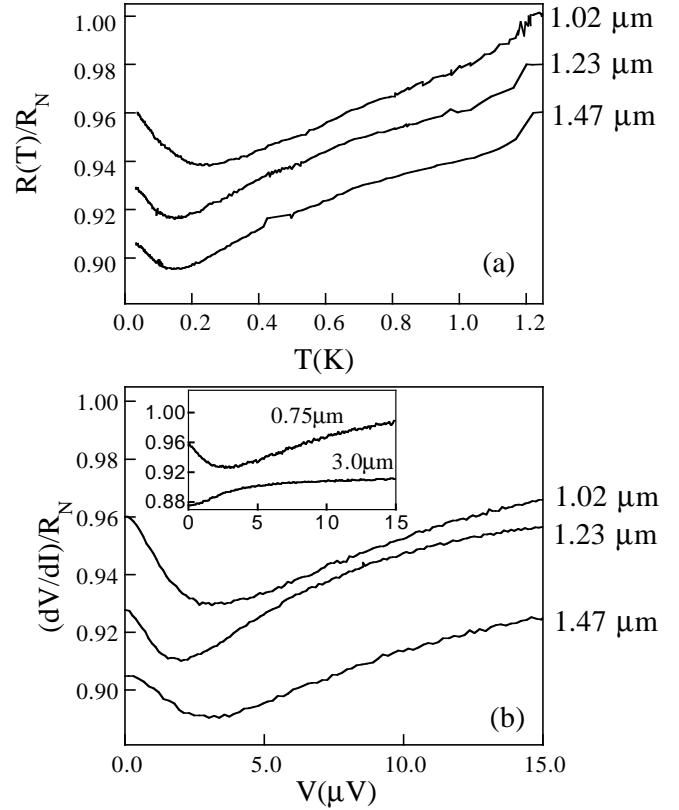 scaled 625}
\end{center}
\caption{(a)  Normalized differential resistance measured as a function
of temperature for wires of length $L$ = 1.02, 1.23, and
1.47 $\mu$m with normal state resistances $R_N$ = 2.41, 2.46,
and 3.69 $\Omega$ respectively.  (b)  Normalized differential resistance  measured as a
function of voltage for the same wires.  The 1.23 and 1.47 $\mu$m curves in (a) and (b) have been offset
by -0.02 and -0.04 $\Omega$ for clarity.  Inset:  Normalized differential resistance for a 0.75 $\mu$m
wire (upper) as a function of dc voltage bias at $T$= 29 mK: the simultaneous measurement of
an adjacent 3.00 $\mu$m wire (lower) acts as a thermometer.}
\label{Fig2}
\end{figure}
inset to Fig. 2(b) shows the differential resistance $R(V)=dV/dI$ of a $L$=0.75 $\mu$m Au wire
from a second sample set as a 
function of dc bias $V$, along with the resistance of
the 3.0 $\mu$m wire immediately adjacent to it.  In the absence of heating
effects, one would not expect the
resistance to change as a function of $V$ across the 0.75 $\mu$m wire, since the
3.0 $\mu$m wire has no dc current flowing through it.  By correlating $R(T)$ of
the 3.0 $\mu$m wire with its change in resistance as a function of $V$ across the
0.75 $\mu$m wire, we can estimate the effective temperature increase as a
function of $V$.  For example, a bias of $V=5$ $\mu$V corresponds to an increase
in the electron temperature in the 3.0 $\mu$m wire to 250 mK from 29 mK.  Since $V_{min}$ is expected to
decrease as
$T$ increases, this increase in the effective temperature would lead to a decrease in
the apparent value of $V_{min}$, as we observe.  In principle, it should be
possible to calculate and correct for this effect by taking into account the
mechanisms for heat generation and dissipation in the sample \cite{Nagaev}; in practice, our
geometry is complex enough to make this very difficult.  Consequently, in the
remainder of the paper, we shall confine ourselves to the discussion of the
temperature dependent resistance $R(T)$.  We should note, however, that our sample
design, with its large normal and superconducting reservoirs, might be expected to
minimize heating due to the dc bias; in samples without such reservoirs (as in
many previous experiments), the problem will be more acute.

We now come to the discussion of the temperature dependent resistance.  The procedure for
calculating the normalized resistance $R(T)/R_N$ from the quasiclassical theory of superconductivity
has been discussed by many authors \cite{Belzig}, and we shall not repeat it in detail here.  Briefly,
one needs to solve the Usadel equation \cite{Usadel}
\begin{equation}
\hbar D_N \frac{\partial ^{2}\theta (\varepsilon,x)}{\partial x^{2}} +
2i\varepsilon \sin
\theta(\varepsilon,x) = 0
\label{eq:eqn1}  
\end{equation}
for the complex angle $\theta$ which parameterizes the nonequilibrium
superconducting Green's functions.  This equation is solved subject to the boundary condition that
\begin{equation}
\theta (\varepsilon )=
\left\{ \begin{array}{ll}
\frac{\pi}{2} + i \frac{1}{2} \ln \frac{ \Delta (T) + \varepsilon
}{\Delta (T)-\varepsilon} & \mbox{if
$\varepsilon <
\Delta (T)$}
\\ i \frac{1}{2} \ln  \frac{\varepsilon +\Delta (T)}{\varepsilon -\Delta (T)} & \mbox{if $\varepsilon
>\Delta (T)$}
\end{array}
\right.
\label{eq:eqn2}  
\end{equation}
at the superconducting reservoir ($x=0$), and $\theta=0$ at the normal reservoir ($x=L$).
The enhanced diffusion coefficient is then obtained by integrating over the
length of the sample
\begin{equation}
D(\varepsilon )= \frac{D_N}{1/L \int_{0}^{L} dx \, \text{sech} ^2 \left[ \Im \theta (\varepsilon , x)
\right]}.
\label{eq:eqn3}  
\end{equation}
Finally, $R(T)/R_N$ is obtained by convoluting $D(\varepsilon )$ with a thermal kernel
\begin{equation}
\frac{R(T)}{R_N}=\left[ \int_{0}^{\infty} \frac{D(\varepsilon )}{2 k_B T \cosh^2 \left(
\frac{\varepsilon}{2k_B T}\right)}  \right] ^{-1}.
\label{eq:eqn4}  
\end{equation}
In their
calculation for this geometry, Nazarov and Stoof \cite{Nazarov} assume that $\varepsilon
\ll \Delta (T)$ at all temperatures $T$; under this assumption, the boundary conditions Eq.(2) at the
superconducting reservoir reduce simply to $\theta = \pi/2$.  A better
approximation can be made by taking into account the temperature dependence of
the gap.  Recently, Petrashov \etal  \cite{Petrashov2} fit $R(T)$ for their Ag/Al
Andreev interferometers to the quasiclassical theory by using $\Delta (T)$ as a
fitting parameter.  The resulting temperature dependence of $\Delta$ they
obtained was in agreement with the predictions of the microscopic theory of
Bardeen, Cooper and Schrieffer (BCS) \cite{Tinkham}; however, the zero temperature value $\Delta
(0)$ was a factor of five smaller than would be expected from $T_c$ of
the Al film.

Almost all analyses of the mesoscopic proximity effect neglect inelastic scattering of the
quasiparticles, assuming that the inelastic scattering length $L_{in}$ is much longer 
\begin{figure}
\begin{center}
\BoxedEPSF{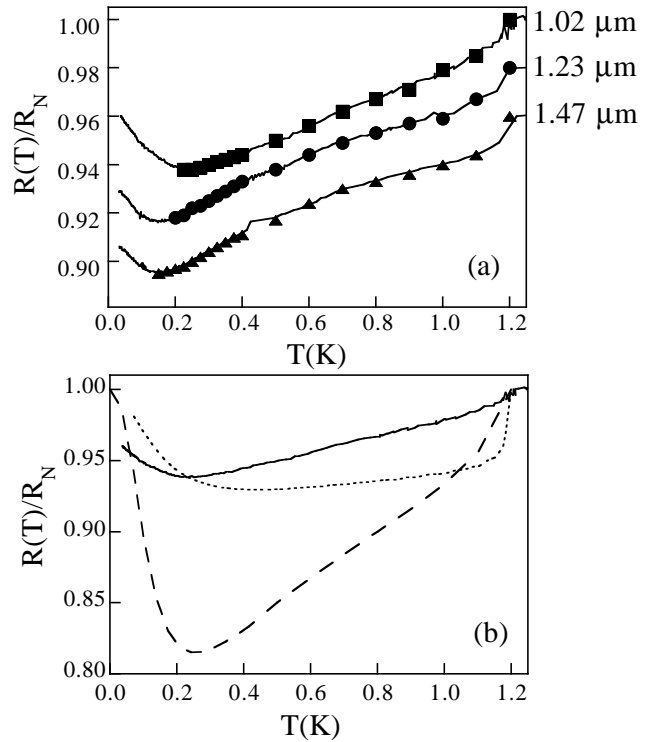 scaled 625}
\end{center}
\caption{(a)  Solid lines: data from Fig. 2(a); solid symbols, simulation of $R(T)/R_{N}$ with a
temperature dependence fitting parameter
$L_{\gamma}$, and a temperature dependent gap, as described in the text.  The 1.23 and 1.47 $\mu$m curves here and in (b) have been offset by
-0.02 and -0.04 $\Omega$ for clarity. (b)  Solid line: data for 1.0 $\mu$m wire from Fig.
2(a); dashed line, simulation of
$R(T)/R_{N}$ for its temperature dependence using the recipe of Petrashov \etal [11], with
$L_{\gamma}$ = 2.0
$\mu$m and $\Delta (0) = \beta \times$
182 $\mu$eV with $\beta$ = 0.2, the same value as used by Petrashov
\etal }
\label{Fig3}
\end{figure}
than the sample
dimensions in the temperature range of interest \cite{lphinote}.  If $L_{in}<L$, $L_{in}$ determines the
effective sample 
length.   Clearly, if $L_{in}$ changes as a function of temperature, this will affect
the measured $R(T)/R_N$.  It is instructive to
investigate the effect of temperature dependent inelastic scattering on the proximity effect.  This can
be accomplished by introducing an imaginary component $\gamma (T)$ to the energy of the quasiparticles
\cite{Yip},
$\varepsilon
\rightarrow \varepsilon + i \gamma (T)$, with a corresponding length $L_{\gamma}
=\sqrt{\hbar D/2\gamma}$.  Figure 3(a) shows the experimental $R(T)/R_N$ curves for the three
wires of Fig. 2, along with fits to the quasiclassical theory using $L_{\gamma}$ as a temperature
dependent fitting parameter.  In order to obtain these fits, we assumed a BCS-like temperature
dependence of the gap, with $\Delta (0)$=182
$\mu$eV estimated from $T_c$ of the Al film.  Below $T_{min}$, the fits become insensitive
to the choice of $\gamma$, due to the fact that $L_{\gamma}$ becomes longer
than the length of the wires.  Consequently, we have only fit the data down to $T_{min}$
\cite{fitnote}.  For comparison, Fig. 3(b) shows
curves calculated assuming a temperature dependent gap with $\Delta (0) = 0.2 \times 182$ $\mu$eV,
but a constant temperature independent
$L_{\gamma}=2.0$ $\mu$m, as in Petrashov \etal \cite{Petrashov2}.  These curves deviate from the
experimental results at almost all temperatures \cite{Petrashovnote}.

\begin{figure}
\begin{center}
\BoxedEPSF{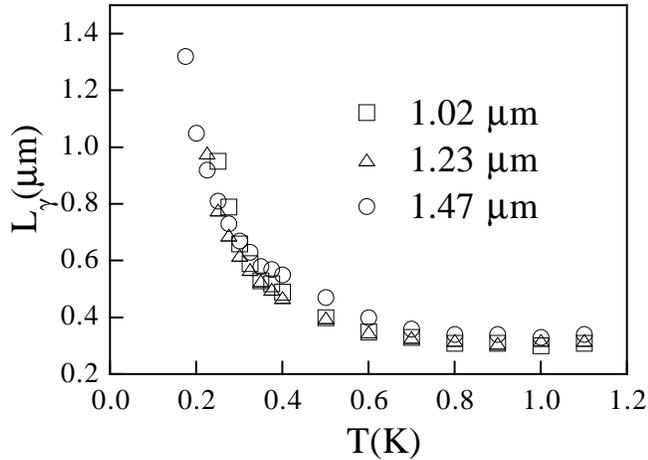 scaled 773}
\end{center}
\caption{$L_{\gamma }(T)$ obtained by fitting to the data in Fig. 2(a) for
the $L$ = 1.02, 1.23, and 1.47 $\mu$m wires.  Even though the
fits were performed independently from one another, the
calculated values lie on the same curve.}
\label{Fig4}
\end{figure} 

Figure 4 shows $L_{\gamma}$ obtained from these fits obtained as described above as a function of $T$ for
the 1.02, 1.24 and 1.47 $\mu$m long wires.  Even though the fits were performed
independently, the resulting $L_{\gamma} (T)$ for each wire is essentially the same. 
$L_{\gamma}$ saturates at higher temperatures at $\sim 0.3$ $\mu$m, and appears to
diverge at lower temperatures.  The values of $L_{\gamma}$ obtained are much smaller than the
experimentally determined value for $L_{\phi} \sim 3.7$ $\mu$m.  In addition, the temperature dependence
of $L_{\gamma}$ is much sharper than the expected temperature dependence of $L_{\phi}$
\cite{Altshuler,Mohanty}.  Consequently, it appears that $L_{\gamma}$ is not directly related to the
phase coherence length in the normal metal.

Recently, Giroud \etal \cite{Giroud} observed a reentrant proximity effect in a Co loop connected
to an Al island, even though $L_{\phi}$ in the Co film was short, as evidenced by the absence of quantum
interference effects.  Our work is another indication that $L_{\phi}$ may not be the most relevant
length scale for the proximity effect.  

We thank C-.J.
Chien, J. Aumentado, and B.W. Alphenaar for valuable discussions.  This work was supported by the
National Science Foundation through DMR-9801892 and by the David and Lucile Packard
Foundation.

\end{multicols}

\begin{references}
\bibitem{samplenote}Each of the two other wires in this sample set were found to have one voltage probe
discontinuous on cooldown, and hence could not be measured.
\bibitem{lphinote}We distinguish here between $L_{in}$ and $L_{\phi}$.  Since, all inelastic events
result in decoherence, but not all phase-breaking mechanisms are inelastic, one usually expects $L_{in}
\ge L_{\phi}$, which does not appear to be the case for these samples.
\bibitem{fitnote}At temperatures below $T_{min}$, the calculated
curves deviate from the experimental values, due to the effect of electron-electron interactions on the
proximity effect \cite{Belzig} which are not taken into account in our simulations.
\bibitem{Petrashovnote}The value of $\Delta (0)$ was chosen to match $V_{min}$.  The overall change in
resistance of the theoretical curve could be reduced to agree with the experimental curve by assuming a
low transparency interface, but the low resistance of our interfaces would argue against this
assumption.
\end{references}
\end{document}